\documentclass[a4paper,UKenglish]{lipics-v2021}
\usepackage{nowidow}
\usepackage{hyperref}
\usepackage{xcolor}
\definecolor{blue}{rgb}{0.25, 0.41, 0.88}
\definecolor{blue}{rgb}{0.0, 0.22, 0.66}
\hypersetup{%
      colorlinks    = true,
  linkcolor = {blue},
   anchorcolor = {blue},
   citecolor = {blue},
   filecolor = {blue},
   urlcolor = {blue},
   menucolor = {blue},
   pagecolor = {blue},
}
\usepackage{booktabs}
\usepackage{siunitx}
\usepackage{multirow}
\usepackage{xspace}
\usepackage[font=rm]{caption}
\usepackage{algorithm}
\usepackage[noend]{algpseudocode}
\usepackage{verbatim}
\usepackage{graphicx}
\usepackage{tabularx}
\usepackage{soul}
\usepackage{natbib}
\newcommand{\method}[1]{{\sf{#1}}}

\newcommand{\var}[1]{\mbox{\emph{#1}}}

\newcommand{\myparagraph}[1]{\paragraph*{#1}}

\usepackage{xcolor}

\newcommand{\aftertabspace}{\vspace*{-2ex}}
\newcommand{\afterfigspace}{\vspace*{-2ex}}

\newcommand{\Lab}[1]{{\mbox{\tt{#1}}}}

\newcommand{\RR}{\method{RR}}
\newcommand{\Prec}{\method{Prec}}
\newcommand{\Succ}{\method{Succ}}
\newcommand{\AP}{\method{AP}}
\newcommand{\NDCG}{\method{NDCG}}
\newcommand{\RBP}{\method{RBP}}
\newcommand{\Robust}{\method{Robust}}
\newcommand{\IPSO}{\method{IPSO}}

\newcommand{\NI}{{\mbox{\tt{ni}}}}
\newcommand{\NS}{{\mbox{\tt{ns}}}}
\newcommand{\EQ}{{\mbox{\tt{==}}}}
\newcommand{\NP}{{\mbox{\tt{**}}}}

\hideLIPIcs

\title{How Much Freedom Does An Effectiveness Metric Really Have?}

\author{Alistair Moffat}{The University of Melbourne, Australia}{ammoffat@unimelb.edu.au}{}{} 
\author{Joel Mackenzie}{The University of Queensland, Australia}{joel.mackenzie@uq.edu.au}{}{} 

\authorrunning{A. Moffat and J. Mackenzie}
\Copyright{A. Moffat and J. Mackenzie}


\category{Author Preprint} 
\relatedversion{}
\acknowledgements{}
\nolinenumbers 

\begin{document}

\maketitle

\begin{abstract}
It is tempting to assume that because effectiveness metrics have free
choice to assign scores to search engine result
pages (SERPs) there must thus be a similar degree of freedom as to
the relative order that SERP pairs can be put into.
In fact that second freedom is, to a considerable degree, illusory.
That's because if one SERP in a
pair has been given a certain score by a metric, fundamental ordering
constraints in many cases then dictate that the score for the second
SERP must be either not less than, or not greater than, the score
assigned to the first SERP.
We refer to these fixed relationships as {\emph{innate pairwise SERP
orderings}}.
Our first goal in this work is to describe and defend those pairwise
SERP relationship constraints, and tabulate their relative
occurrence via both exhaustive and empirical experimentation.

We then consider how to employ such innate pairwise relationships in
IR experiments, leading to a proposal for a new measurement paradigm.
Specifically, we argue that tables of results in which many different
metrics are listed for champion versus challenger system comparisons
should be avoided; and that instead a single metric be argued for in
principled terms, with any relationships identified by that metric
then reinforced via an assessment of the innate relationship as to
whether other metrics are likely to yield the same system-vs-system
outcome.
\end{abstract}
 
\section{Introduction}

Information retrieval mechanisms are often compared using
{\emph{offline evaluation}} techniques, also sometimes known as
{\emph{batch evaluation}}.
In such a comparison a corpus of suitable documents is acquired, a
set of representative information needs ({\emph{topics}}) and
corresponding {\emph{queries}} (one or more per topic) is developed,
and relevance judgments connecting the information needs and the
topics are solicited.
A comparison of two IR systems -- perhaps as a {\emph{champion}}
versus {\emph{challenger}} experiment -- can then be carried out by
executing the set of queries using each of the two systems, to
generate pairs of {\emph{search engine result pages}} (or
{\emph{SERPs}}), and then comparing the quality of those pages, with
quality assessed via the {\emph{usefulness}} of that SERP in terms of
addressing the corresponding topic's information need.
This work-flow relies on the use of one or more {\emph{effectiveness
metrics}}, each of which is a categorical-to-numeric mapping that
takes a SERP and the relevance judgments for that topic as inputs,
and returns a real-valued score.
Those numeric scores are regarded as being surrogates that quantify
the SERP's usefulness to the hypothetical user, or to a conjectured
community of similar users.
The systems are then compared based on their paired SERP scores, with
one such comparison being performed for each of the selected metrics.
{\citet{s10-fntir}} describes this process in more detail.

A large number of effectiveness metrics have been proposed in the
literature, with more added each year.
As one simple example, {\emph{reciprocal rank}} (\RR) scores
binary-valued SERPs according to the inverse of the rank of their
first relevant document, assuming that document relevance is a binary
indicator.
Result pages that have a relevant document in the first position of
the SERP are assigned a score of $1.0$; SERPs with the first relevant
document at rank two get a score of $0.5$; and SERPs that lack even a
single relevant document (down to some depth $k$) are assigned a
score of zero.

Each such proposal for a metric is (or at least, should be) motivated
by a corresponding {\emph{user model}} that is argued as representing
the behavior of the community of users of the search system in
question, and hence reflecting the way in those users acquire
usefulness {\citep{mbst17acmtois}}.
Those models usually assume that the user peruses the elements in the
SERP from the top down, in the same order as they are presented.
In this framework {\RR} models the behavior of users who seek a
single relevant document, and who assess the usefulness of each SERP
through the ``reward to effort'' ratio, with effort captured by the
number of SERP elements considered, be they snippets/captions, or
whole documents.
{\citet{mbst17acmtois,sigir22mmta}} discuss user models and they way
that they are connected to effectiveness metrics.

System-vs-system comparisons -- as appear in many IR papers,
including in this journal -- then typically present results for
multiple effectiveness metrics, thereby ``covering the field'' and
``hedging their bets''.
While including multiple metrics in a
system comparison is certainly one way of adding generality, such
approaches will always be vulnerable to criticism in connection with
the exact choice of metrics used.
If the community of users being discussed is believed to engage in
patient or deep retrieval tasks and the primary measurement device is
thus a deep metric, should the second ``validation'' metric be
another deep metric?
Or should the next metric employed be a shallow one, to provide a
more illustrative complementary evaluation?

Our work here brings a 
new perspective to this typical experimental pipeline.
Instead of seeking significance across each of a
palette of specific metrics, we propose that researchers employ a
single metric, namely the one that best suits the task and community
of users that they wish to serve.
That is, we argue that a single appropriate metric be reported,
matching the user model that best fits that community's anticipated
aggregate search behavior.
Then, rather than adding further hedging metrics to their
evaluations, we propose that researchers employ innate SERP versus
SERP relationships to support their claims in regard to generality.
We define exactly what is meant by that shortly; for now, we
summarize the idea as being a relationship between two SERPs that can
in some cases result in their relative score orderings being known
for {\emph{any and every effectiveness metric}} when evaluated to
some given depth limit of~$k$.

That is, as a system-vs-system corroboration tool, we seek out
pairwise SERP relativities that must be valid for {\emph{all
reasonable metrics}}.
Such pairings are, for typical values of $k$, surprisingly frequent;
and when they arise consistently across a set of topics they provide
an indication that the relative system orderings observed using the
metric of choice would also be likely to be detected by other
metrics.
This ``universal hedge'' role then allows the primary metric to be
chosen on a principled basis to suit the community of users that is
being modeled, without multiple other metrics being required as part
of the presentation.
Given that context, the path through this paper considers the
following research questions:
\smallskip

\noindent{\bf{RQ1}}
Are there fundamental relationships between SERPs that can
allow the ordering of the effectiveness scores of two SERPs to be
known independently of the actual effectiveness metric used?
\smallskip

\noindent{\bf{RQ2}}
Do such relationships, if they exist, occur sufficiently frequently
as to be informative?
\smallskip

\noindent{\bf{RQ3}}
To what extent do any derived innate relationships agree with
system-vs-system evaluations carried out
in the traditional way?
\smallskip

\noindent{\bf{RQ4}}
Can those innate relationships, if they exist, be used in a way that
adds validity to the measured outcome of an experiment?
\medskip

\noindent
The next section summarizes the principles that underly IR
effectiveness metrics, and introduces the required ideas for our
proposed approach.
 \section{Innate Pairwise SERP Orderings}

We now consider ways in which SERP pairs can be regarded as being
innately ordered by virtue of fundamental relationships ({\bf{RQ1}}).
We starts with definitions, then introduce two ordering rules and
argue that they are universal to SERP-based IR evaluations.

\myparagraph{Definitions}

A SERP can be thought of as being an ordered $k$-vector,
$\mathbf{r}=[r_i \mid 1\le i \le k]$, where $r_i$ is the relevance
grade associated with the SERP's $i$\,th item.
The $r_i$ values can be arbitrary (graded relevance), but are often
binary, $r_i\in\{\Lab{0},\Lab{1}\}$, the case assumed here.
A SERP is thus a relevance-mapped $k$-prefix of a complete ordering
of the $n$ documents in the IR system.
We assume that SERPs can range from being all-$\Lab{0}$ to
all-$\Lab{1}$, and in the first instance focus on some single~topic.

In terms of measurement, the $r_i$ values arise on either an ordinal
scale (relevance {\emph{grades}}) or a ratio scale (relevance
{\emph{gains}}).
But the SERPs themselves are categorical data, since there is no
ordering that can be applied to every possible pair of SERPs.
For example, it is completely unclear whether the SERP {\tt{[1,0,0]}}
should be better than or worse than {\tt{[0,1,1]}} in terms of its
benefit to a user searching for information.
Users might prefer either.

Nevertheless, some SERP relativities can be inferred from the fact
that the relevance values $r_i$ are (at a minimum) on an ordinal
scale.
For example, the SERP {\Lab{[1,1,0]}} cannot be inferior to the item
SERP {\Lab{[1,0,0]}}, because the latter has no rank positions at
which it exceeds the former.
Similarly, provided that SERPs are assumed to be consumed from
left-to-right (as shown here, corresponding to top to bottom
consumption on a screen or results page in the more usual
presentation), the SERP {\Lab{[1,1,0]}} cannot be inferior to
{\Lab{[1,0,1]}}, because the former has the same total relevance, and
that relevance occurs at earlier rank positions.

\myparagraph{Fundamental Ordering Rules}

The example SERP pairings discussed in the previous paragraph are
instances of two well-known relationships that allow a partial
ordering of SERPs to be derived, with the presentation and
terminology used here taken from {\citet{ieeeaccess22moffat}}:
\begin{itemize}

\item{\emph{Rule 1}:}
SERP {\Lab{S1}} is {\emph{non-inferior}} to SERP {\Lab{S2}}, written
$\Lab{S1}\succeq\Lab{S2}$, if every
element of {\Lab{S1}} is greater than or equal to the corresponding
element of {\Lab{S2}} in terms of their ordinal document relevance
labels;

\item{\emph{Rule 2}:}
SERP {\Lab{S1}} is also {\emph{non-inferior}} to SERP {\Lab{S2}} if
{\Lab{S2}} can be formed as a transformation of {\Lab{S1}} in which
one or more of {\Lab{S1}}'s elements are swapped rightwards and
exchanged with elements of strictly lower document relevance that
move leftward.
\end{itemize}

\noindent
Rule~1 is an absolute relationship that does not rely on the
documents in the SERP being examined in a top-down manner.
It ensures that any direct SERP-to-SERP relationships that are a
result of weakening one or more of
the individual relevance values
are reflected as whole-of-SERP relationships, for example,
${\Lab{[1,1,0]}}\succeq{\Lab{[1,0,0]}}$.

Rule~2 then captures the SERP-to-SERP relationships that can be added
if it is assumed that the SERP is examined sequentially from
top-to-bottom (here, left-to-right).
It is this rule that asserts
${\Lab{S1}}={\Lab{[1,1,0]}}\succeq{\Lab{[1,0,1]}}={\Lab{S2}}$,
because the second {\Lab{1}} in {\Lab{S1}} has swapped rightward, and
a relevance value of lower grade has moved leftward, in order to form
{\Lab{S2}}.
Of course, when relevance is binary there are only two grades
available, {\Lab{0}} and {\Lab{1}}.

One important point to note is that while both of these rules imply
that if ${\Lab{S1}}\succeq{\Lab{S2}}$ then {\Lab{S2}} precedes
{\Lab{S1}} when the two SERPs are considered to be vectors and
compared lexicographically, the converse is {\emph{not}} true,
because lexicographic sorting results in a total order.
Note also that the two rules describe non-inferiority, $\succeq$, and
not superiority, $\succ$.
For example, it might be tempting to claim that
$\Lab{[1,1,0]}\succ\Lab{[1,0,0]}$ and is strictly superior.
But we do not wish to require that any given user inspects all $k$
documents (indeed, the metric {\RR} provides an immediate
counter-example), and hence the best that can be said is that
${\Lab{[1,1,0]}}\succeq{\Lab{[1,0,0]}}$.

If {\Lab{S1}} is non-inferior to {\Lab{S2}}, then {\Lab{S2}} is
{\emph{non-superior}} to {\Lab{S1}}, written
$\Lab{S2}\preceq\Lab{S1}$.
Two further options then arise:
\begin{itemize}

\item{\emph{Equality}:} If two $k$-SERPs {\Lab{S1}} and {\Lab{S2}}
are such that {\Lab{S1}} is non-inferior to {\Lab{S2}}
($\Lab{S1}\succeq\Lab{S2}$) and {\Lab{S2}} is non-inferior to
{\Lab{S1}} ($\Lab{S2}\succeq\Lab{S1}$) then they are
{\emph{equal}}, denoted as $\Lab{S1}=\Lab{S2}$.

\item{\emph{Non-Separability}:} If two $k$-SERPs {\Lab{S1}} and
{\Lab{S2}} are such that {\Lab{S1}} is {\emph{not}} non-inferior to
{\Lab{S2}} (that is, $\Lab{S1}\not\succeq\Lab{S2}$), and {\Lab{S2}}
is also {\emph{not}} non-inferior to {\Lab{S1}} (that is,
$\Lab{S2}\not\succeq\Lab{S1}$), then they cannot be assigned a
relativity within the innate requirements of Rule~1 and
Rule~2, and are denoted as being {\emph{non-separable}}.
\end{itemize}

\smallskip
\noindent
Any pair of $k$-SERPs $\Lab{S1}$ and
$\Lab{S2}$ can thus be categorized as being equal, non-separable, or
{\emph{separable}}, with that third category covering the cases when
$\Lab{S1}\not=\Lab{S2}$ and either $\Lab{S1}\succeq\Lab{S2}$ or
$\Lab{S1}\preceq\Lab{S2}$.

\begin{figure}[t!]
\centering
\includegraphics[page=1,width=100mm,trim=70mm 70mm 80mm 25mm,clip]{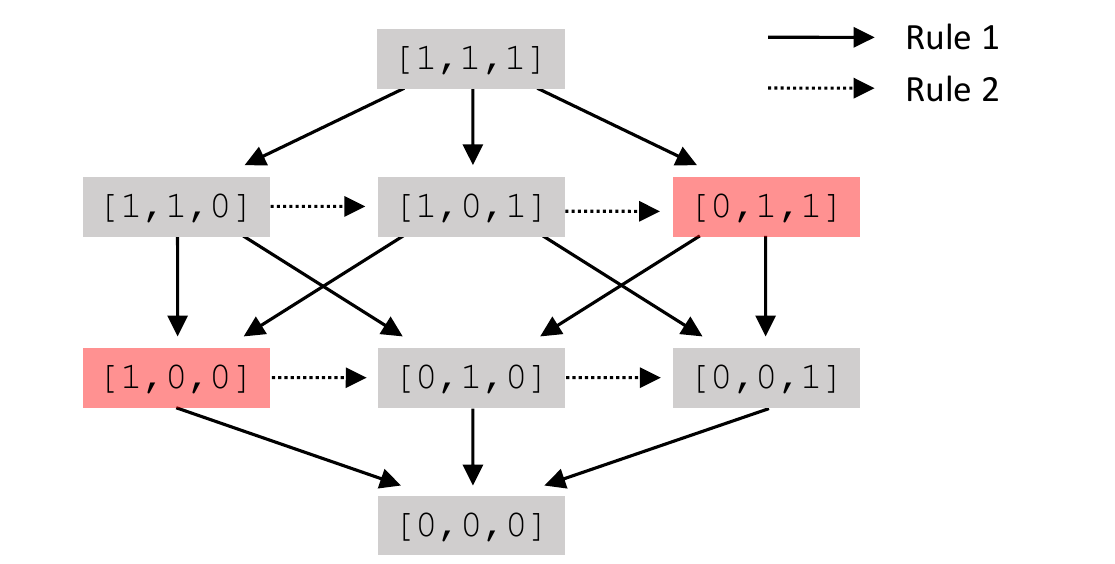}
\caption{Hasse diagram illustrating the innate non-inferiority
relationships amongst SERPs of length $k=3$.
\label{fig-hasse}}
\end{figure}

\myparagraph{Exhaustive Enumeration}

Figure~\ref{fig-hasse} summarizes all pairwise relationships amongst
the $2^k=8$ SERPs of length $k=3$.
Each arrow indicates a $\succeq$ relationship between the two
indicated SERPs; and arrows are omitted in cases where they can be
inferred via transitivity, with directed paths having the usual
interpretation.
There is no directed path between the two highlighted elements
{\Lab{[1,0,0]}} and {\Lab{[0,1,1]}} and hence no non-inferiority
relationship that connects them.
As was noted earlier, that SERP pair is non-separable.

\begin{figure}[t!]
\centering
\includegraphics[width=50mm,trim=10mm 250mm 160mm 10mm,clip]{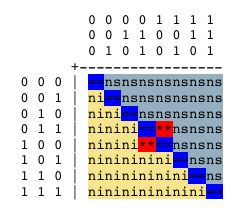}
\caption{All SERP pairs of length $k=3$, and the relationships
between them.
Red cells indicate non-separability.
\label{fig-rbp3}}
\end{figure}

Figure~\ref{fig-rbp3} shows the same universe of eight $k=3$ SERPs
using a different presentation.
In each cell the SERP common to that row (with SERP bits read
left-to-right, and $r_1$ first) is compared to the SERP common to the
corresponding column (with SERP bits top-to-bottom, and
$r_1$ at the top).
The cells are color-coded: dark blue for SERPs that are equal
(denoted {\EQ}); yellow for non-inferior relationships ($\succeq$,
denoted {\NI}); light blue for non-superior ($\preceq$, denoted
{\NS}); and red for non-separable (marked in the figure using {\NP}).
This grid thus captures all $64$ relationships possible in
Figure~\ref{fig-hasse}, and makes clear that there is only a single
pair of $k=3$ SERPs that are non-separable, the pair highlighted in
Figure~\ref{fig-hasse}.

\myparagraph{Freedom!}

We reiterate that while the score relativities shown in
Figure~\ref{fig-rbp3} are innate and that there is thus little scope
for divergence from the defined pairwise orderings, there are,
nevertheless, many possible effectiveness metrics, even when $k=3$.
That is because there is an infinity of score {\emph{values}} that
can be assigned that are compliant with the required pairwise SERP
ordering constraints.
For example, ${\Prec}@3$ assigns a different effectiveness score to
each row of SERPs in Figure~\ref{fig-hasse}, has four different
values possible ($0$, $1/3$, $2/3$, and $1$), and gives
$\Lab{[1,0,1]}$ and $\Lab{[0,1,1]}$ the same score; whereas $\RR$@$3$
also assigns four values across the eight SERPs, but they are
different ones ($0$, $1/3$, $1/2$, and $1$), and $\RR$@$3$ gives
those same two SERPs different scores.
Similarly, $\RBP(0.5)@3$ ({\emph{rank-biased precision}}, see
{\citet{mz08-tois}} for a description) maps $k=3$ SERPs to a total of
eight different scores, with each of the $2^k$ possible SERPs
receiving a different metric score.
Nevertheless, two SERPs related by $\succeq$ must have metric scores
that numerically also comply; and they must comply in every plausible
metric.

\begin{table}[t]
\caption{The three possible effectiveness metric score
relationships for the single non-separable SERP pair that arises when
$k=3$.
\label{tbl-k3-metrics}}
\begin{center}
\renewcommand{\tabcolsep}{2em}
\begin{tabular}{ll}
\toprule
Ordering
	& Metrics $M(\cdot)$ with that ordering
\\
\midrule
{$M(\Lab{[1,0,0]}) < M(\Lab{[0,1,1]})$}
	& ${\Prec}@3$, $\RBP(0.8)@3$, ${\AP}@3$, ${\NDCG}@3$
\\[0.5ex]
{$M(\Lab{[1,0,0]}) = M(\Lab{[0,1,1]})$}
	& ${\Succ}@3$, ${\RBP}((\sqrt{5}-1)/2)@3$
\\[0.5ex]
{$M(\Lab{[1,0,0]}) > M(\Lab{[0,1,1]})$}
	& ${\RR}@3$, ${\RBP}(0.5)@3$
\\
\bottomrule
\end{tabular}
 \end{center}
\end{table}

It is only for non-separable pairs that the metric has ordering
freedom.
Table~\ref{tbl-k3-metrics} provides examples of the three ordering
relationships possible when considering the single $k=3$
non-separable pair shown in Figures~\ref{fig-hasse}
and~\ref{fig-rbp3}.
The metric ${\Succ}@k$ is $1$ if there is a relevant document anywhere
in the top $k$, and $0$ otherwise; {\AP} is {\emph{average
precision}} {\citep{bv05trecbook}}; and {\NDCG} is {\emph{normalized
discounted cumulative gain}} {\citep{jk02-tois}}.
Note that in the case of parameterized metrics like $\RBP$ the number
of different scores produced and the numeric orderings attached to
non-separable pairs can be affected by the metric parameter.
In particular, the use of the golden ratio
$\phi=(\sqrt{5}-1)/2\approx 0.62$ in the middle row of
Table~\ref{tbl-k3-metrics} gives equality for the non-separable SERP
pair and thus allows $\RBP$ to yield all three possible arrangements
of the two SERPs making up that non-separable pair.

\myparagraph{Other Score Orderings}

The choice of row and column ordering in Figure~\ref{fig-rbp3} is
arbitrary (after all, SERPs are categorical data), and the
visualization remains valid if the rows are permuted, or the columns
are permuted, or both.
There will always be eight blue cells, two red cells, and so on.
That is, while Figure~\ref{fig-rbp3} employs a row and column
ordering based upon lexicographic SERP representations as
multi-dimensional vectors, other orderings could also be applied.
As it turns out, a lexicographic ordering corresponds to the SERP
score ordering induced by $\RBP(0.5)$.
That is, Figure~\ref{fig-rbp3} can also be viewed as indicating the
locations in the {\RBP}$(0.5)$ score spectrum at which the
non-separabilities arise when {\RBP}$(0.5)$ is compared against other
metrics.
The blue cells form a perfect diagonal because the row and column
orderings are identical.

\begin{figure}[t!]
\centering
\renewcommand{\tabcolsep}{1em}
\begin{tabular}{@{}cc@{}}
\includegraphics[width=65mm,trim=10mm 110mm 5mm 0mm,clip]{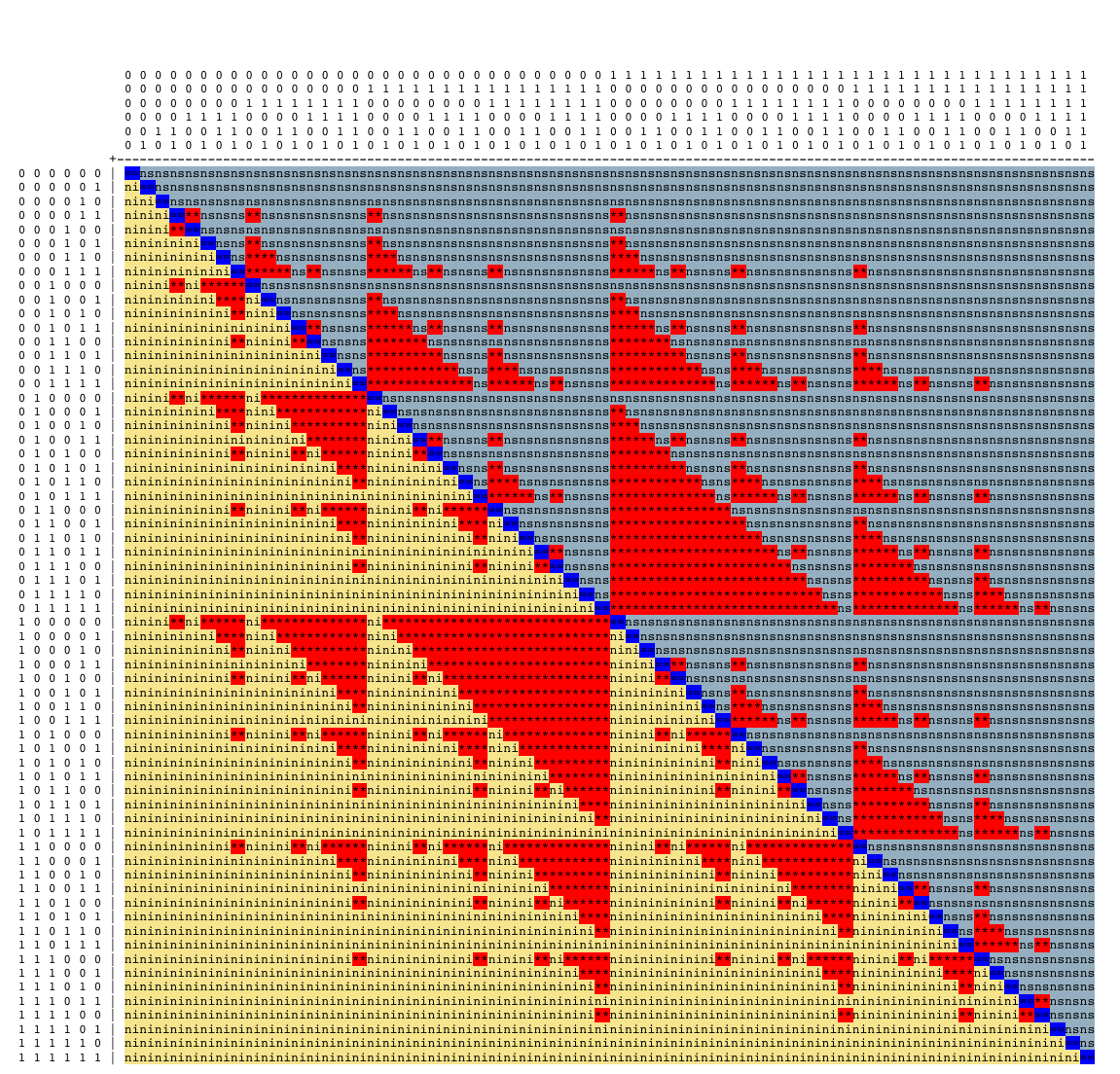}
&
\includegraphics[width=65mm,trim=10mm 110mm 5mm 0mm,clip]{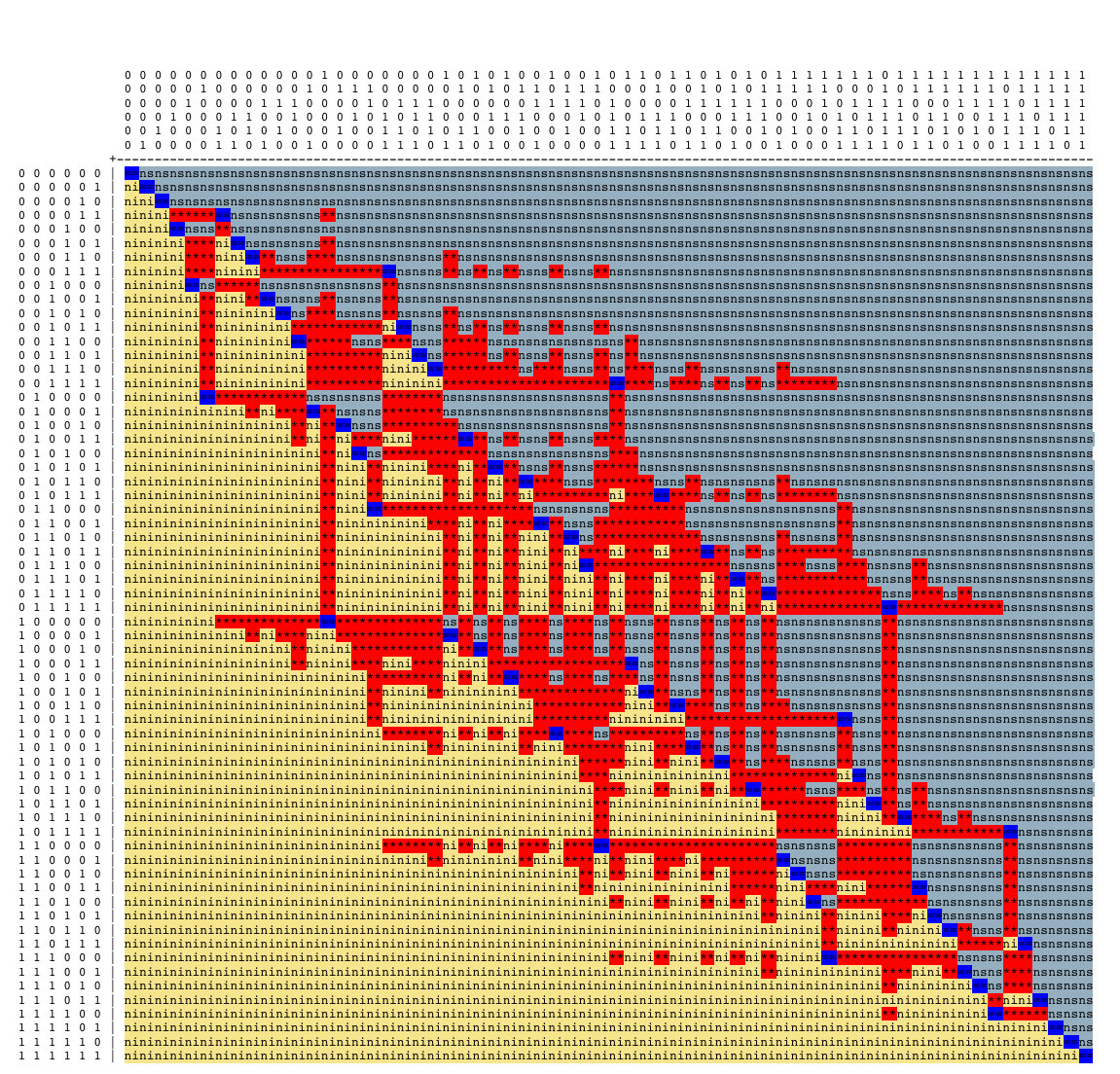}
\\
(a) {\RBP}$(0.5)$ (lexicographic) on both
&
(b) {\RBP}$(0.5)$ on left, {\NDCG} on top
\\
\includegraphics[width=65mm,trim=10mm 110mm 5mm 0mm,clip]{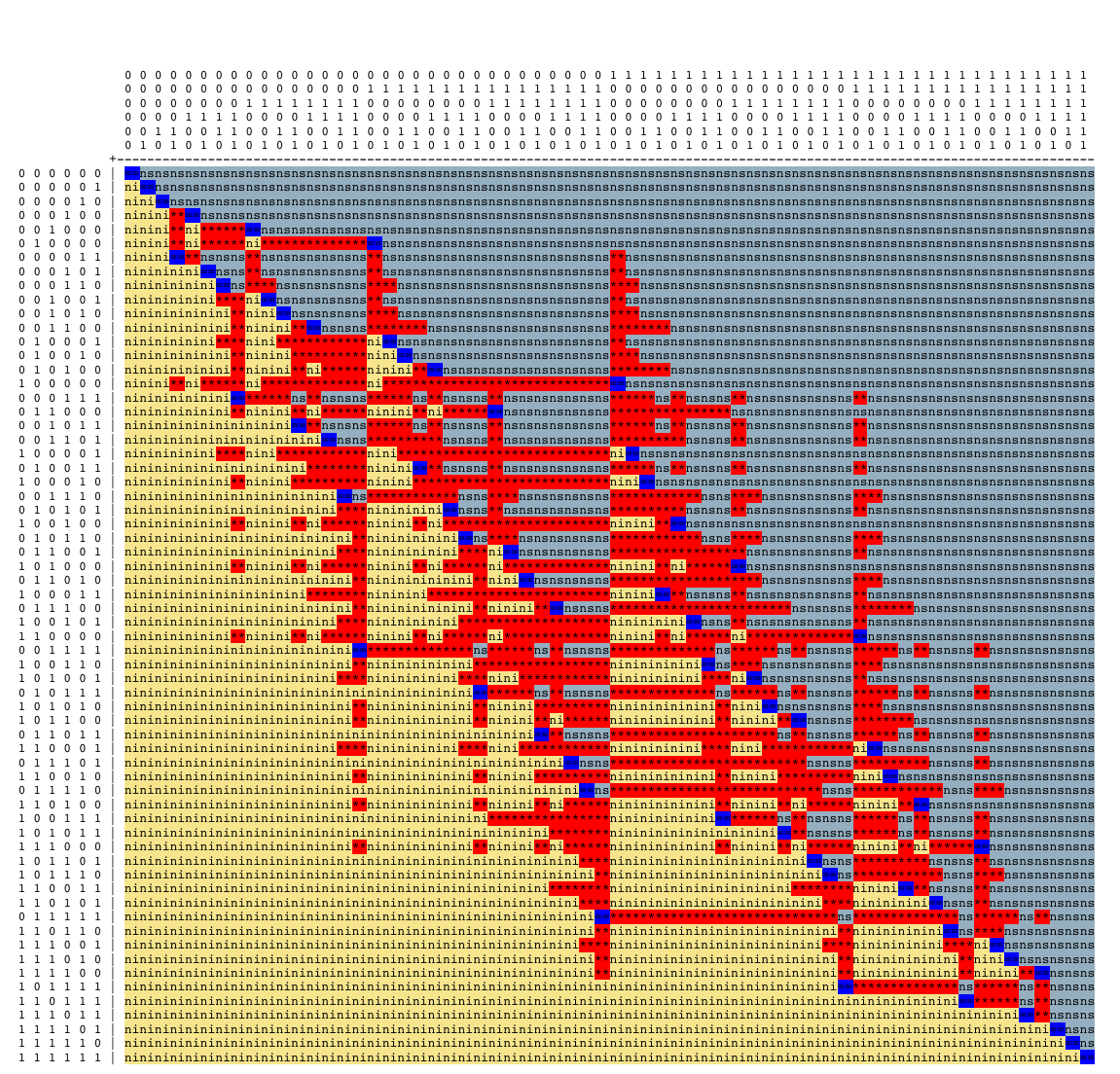}
&
\includegraphics[width=65mm,trim=10mm 110mm 5mm 0mm,clip]{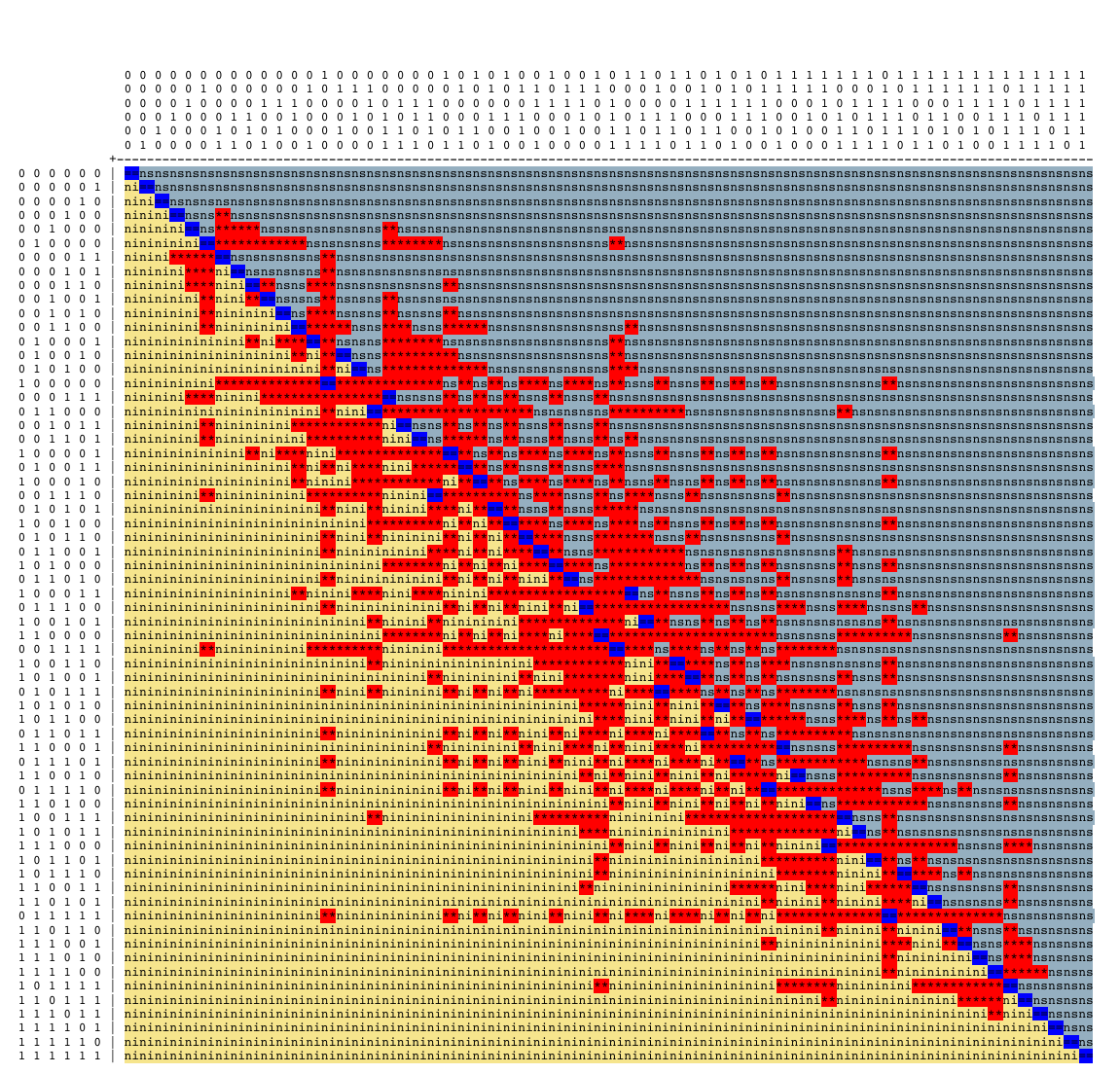}
\\
(c) {\AP} on left, {\RBP}$(0.5)$ on top
&
(d) {\AP} on left, {\NDCG} on top
\\
\end{tabular}
\caption{All SERP pairs of length $k=6$, plotted using the same
colors as in Figure~\ref{fig-rbp3}, with each grid consisting of
$2^6\times 2^6=4{,}096$ cells.
Each pane has exactly the same number of red cells; they indicate
SERP pairs where the two metrics are {\emph{permitted}} to disagree
on the relative ordering of the two SERPs, and are not in any way an
indication that they {\emph{do}} disagree.
\label{fig-k6}}
\end{figure}

Figure~\ref{fig-k6} shows what happens when SERPs of length $k=6$ are
categorized in the same manner.\footnote{We acknowledge that the axis
labels are not readable at A4-print scale, and require ``zoom in'' on
the soft-copy to be viewed.
However our purpose here is to show the overall patterns that emerge,
and not for individual SERP pairs to be identifiable via their
labels.
The reader is invited to think of this figure as being ``art'', not
``science''.}
Pane~(a) directly corresponds to Figure~\ref{fig-rbp3}, and orders
the SERPs on both axes lexicographically, that is, by their
{\RBP}$(0.5)$ scores.
Within pane~(a) Figure~\ref{fig-rbp3} repeats eight times down the
diagonal; the corresponding $k=4$ pattern appears four times; and the
$k=5$ pattern arises twice.
There are several other symmetries and motifs that recur in a
fractal manner.

In Figure~\ref{fig-k6}(b), the rows are still ordered by
{\RBP}$(0.5)$, but the columns are permuted into increasing {\NDCG}
score order {\citep{jk02-tois}}.
Exactly the same number of each color cell is present, but they are
now in a different arrangement.
In pane~(c) the rows are ordered by {\AP} {\citep{bv05trecbook}} and
the columns are ordered by {\RBP}$(0.5)$, yielding a third
presentation of the same underlying data; and then pane (d) completes
the set, with the rows ordered by {\AP} and the columns ordered by
{\NDCG}.
Across the four panes the three sets of $64$ blue cells have
Kendall's correlations, in the context of the four corresponding axis
orderings, of $\tau=1.000$, $\tau=0.796$, $\tau=0.786$, and
$\tau=0.976$ respectively.
Not unexpectedly, {\AP} is more closely correlated with {\NDCG} than
is {\RBP}$(0.5)$, even when $k$ is just~six.

\myparagraph{Computing SERP Pair Categorizations}

\begin{algorithm}[t]
\begin{algorithmic}[2]
\State $\var{cumul} \leftarrow 0$
\State $\var{been\_neg} \leftarrow \var{been\_pos} \leftarrow {\mbox{\bf{false}}}$
\For{$i \leftarrow 1$ {\bf{to}} $k$}
  \State $\var{cumul} \leftarrow
  	\var{cumul} + \Lab{S1}[i]$
\State $\var{cumul} \leftarrow \var{cumul} - \Lab{S2}[i]$
  \If{$\var{cumul} < 0$}
    \State $\var{been\_neg} \leftarrow {\mbox{\bf{true}}}$
  \EndIf
  \If{$\var{cumul} > 0$}
    \State $\var{been\_pos} \leftarrow {\mbox{\bf{true}}}$
  \EndIf
\EndFor
\If{$\var{been\_neg}$ {\bf{and}} $\var{been\_pos}$}
  \State \Return {\sf{non-separable}}
\ElsIf{$\var{been\_pos}$}
  \State \Return {\sf{non-inferior}}
\ElsIf{$\var{been\_neg}$}
  \State \Return {\sf{non-superior}}
\Else
  \State \Return {\sf{equal}}
\EndIf
\end{algorithmic}
 \caption{Comparing SERPs to establish their innate ordering
relationship.
Each of $\Lab{S1}$ and $\Lab{S2}$ is a $k$-vector of $\Lab{0}$s and
$\Lab{1}$s.
\label{alg-compare}}
\end{algorithm}

Algorithm~\ref{alg-compare} describes the comparison process used to
compare pairs of SERPs and thus allow the creation of
Figures~\ref{fig-rbp3} and~\ref{fig-k6}.
A linear scan tracking the cumulative sum of their element-by-element
differences is sufficient to infer the relationship between any pair
of $k$-SERPs {\Lab{S1}} and {\Lab{S2}}.
In particular, if there is no depth prior to $k$ at which {\Lab{S1}}
has fewer one-bits than has {\Lab{S2}}, then {\Lab{S1}} must be
either equal to {\Lab{S2}}, or non-inferior to it.
On the other hand, if {\Lab{S1}} has had both strictly fewer one-bits
than {\Lab{S2}} at some point $k'\le k$, and also strictly more one-bits
than {\Lab{S2}} at some other point $k''\le k$, then {\Lab{S1}} and
{\Lab{S2}} are non-separable.

\begin{figure}[t!]
\centering
\includegraphics[width=85mm,trim=25mm 60mm 125mm 30mm,clip]{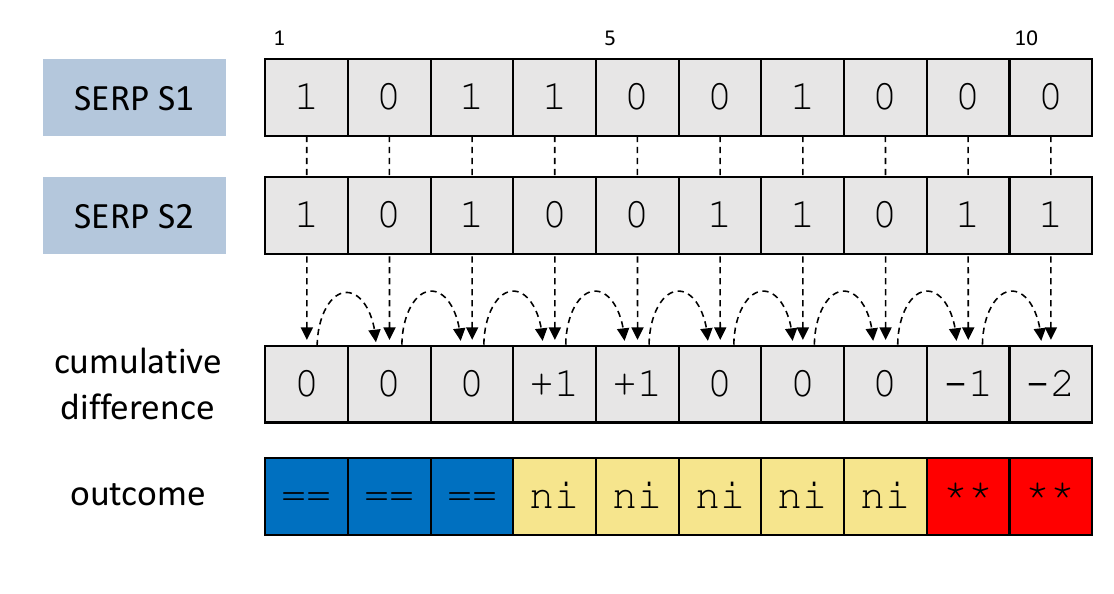}
\caption{Comparing SERPs to get an innate pairwise ordering.
In this example, either {\Lab{S1}} or {\Lab{S2}} might be assigned
the higher score for metrics computed to depth $k = 10$, possible
because the two SERPs are non-separable.
But at depths $k=4$ through to $k=8$ SERP {\Lab{S1}} cannot be
assigned a lower effectiveness score than SERP {\Lab{S2}}.
\label{fig-compare}}
\end{figure}

Figure~\ref{fig-compare} illustrates the application of
Algorithm~\ref{alg-compare}.
At depths $i=1$, $i=2$, and $i=3$ the number of one-bits remains
balanced and they appear in the same positions, and so the two SERPs
are judged to be equal.
From depth $i=4$ the cumulative difference is positive, and so the
outcome switches to ``non-inferior''.
Finally, at $i=9$ the cumulative sum enters negative territory, and
from that point onward the two SERPs must be regarded as being
non-separable.
Hence, any particular metric is free to assign any of the three
possible relative orderings (Table~\ref{tbl-k3-metrics}) to these two
$10$-SERPs.
On the other hand, if only the first $k=5$ elements of $\Lab{S1}$ and
$\Lab{S2}$ are considered, then {\emph{every}} metric must assign a
score to $\Lab{S1}$ that is not lower than the score assigned to
$\Lab{S2}$.
This variation with $i$ is typical: at first, any pair of SERPs are
equal (when $i=0$); and then, after some number of relevance values
have been considered, one of them gains the upper hand.
That advantage either remains until the target depth $k$ is reached,
or until the other SERP in turn claims the majority of one bits.
If the latter occurs at any point the non-separability outcome cannot
be subsequently reversed, and continues to hold as $i$ is further
increased.
No combination of further bits added in Figure~\ref{fig-compare} --
positions $11$ to $20$, for example -- could alter the outcome that
results once $k=9$ is reached.
Of course, each metric assigns a score to each of $\Lab{S1}$ and
$\Lab{S2}$ for each value of $k$, and so for any given metric
there is a score relativity.
Once $k\ge 9$ that relativity can be in either direction.

We can now answer {\bf{RQ1}} in the affirmative: there are indeed
sometimes fundamental relationships between SERPs that dictate the
ordering of ``@$k$'' effectiveness scores, independently of the
actual effectiveness metric used, and irrespective of the
effectiveness scores assigned by that metric.
We next explore implications of that in the context of
typical IR batch evaluation scenarios, leading to a new way of
presenting IR effectiveness results.
 \section{A New Perspective -- IPSO}

We are now in a position to consider {\bf{RQ2}}, {\bf{RQ3}}, and
{\bf{RQ4}}.
First a test environment is described, and then using that data we
explore the implications of these {\emph{innate pairwise SERP
orderings}}.
For brevity -- and because we think it is a fitting acronym -- we
reduce ``innate pairwise SERP ordering'' to {\IPSO}.
We then present our proposal for a new way of presenting IR
system-vs-system experimental results, using {\IPSO} as an additional
indicator.

\myparagraph{Experimental Setup}

We make use of the runs submitted to the 2004 TREC {\Robust} track
{\cite{v04trec}}.
A total of $110$ systems were represented via runs, with each system
responding to a total of $249$ queries on documents from TREC Disks 4
and 5, excluding the Congressional Record from Disk 4.
The official evaluation metric was {\AP}, with ${\Prec}@10$ also
reported.
The {\Robust} relevance judgments contain an average of $1250.6$
judgments per topic, of which $69.9$ had relevance grades of $1$ or
greater.
Our experiments assume binary judgments, with any documents judged
$\ge1$ taken to be relevant, and all unjudged documents assumed to be
non-relevant.

\myparagraph{Exhaustive Enumeration}

RQ2 asks how frequently SERP pairs can be expected to display the
innate ordering characteristics captured by the $\succeq$
relationship.
We explore that in two ways -- first via exhaustive enumeration of
SERP pairs, and then by considering the SERP pairs that arise in a
typical IR experimental setting.

\begin{table}[t]
\centering
\caption{Fraction of SERP pairs of length $k$ that are equal,
separable, and non-separable, based on enumeration ($k\le15$) and
random generation of a billion SERP pairs ($k\ge20$).
\label{tbl-allpairs}}
\newcommand{\tabent}[1]{\makebox[12mm][c]{#1}}
\renewcommand{\tabcolsep}{3mm}
\begin{tabular}{cccc}
\toprule
$k$
	& \tabent{\sf{equal}}
		& \tabent{\sf{separable}}
			& \tabent{\sf{non-sep.}}
\\
\midrule
5
	& 3.12\%
		& 83.98\%
			& 12.89\%
\\
10
	& 0.10\%
		& 67.08\%
			& 32.81\%
\\
15
	& 0.00\%
		& 55.97\%
			& 44.02\%
\\
20
	& 0.00\%
		& 48.91\%
			& 51.09\%
\\
50
	& 0.00\%
		& 31.43\%
			& 68.57\%
\\
100
	& 0.00\%
		& 22.34\%
			& 77.66\%
\\
\bottomrule
\end{tabular}
 \aftertabspace
\end{table}

Table~\ref{tbl-allpairs} categorizes the SERP pairs as $k$ increases,
combining non-inferior and non-superior into a single separable
grouping, as was described above.
To form the table the values for $k\le15$ were computed exactly, via
exhaustive counting over all possible $2^{2k}$ SERP-pair combinations
using binary ($\Lab{0}$ or $\Lab{1}$) relevance judgments.
The three values for $k\ge20$ are estimates, based on assessment of
$10^9$ randomly-generated SERP pairs of each length, where random
means that $\Lab{0}$s and $\Lab{1}$s are equally likely at each rank
from $1$ through to~$k$ in each of the SERPs.

The table indicates that a useful fraction of all possible SERP pairs
can be innately ordered for moderate evaluation depths.
For example, when $k=10$, fully two-thirds of all SERP pairs have a
fixed relationship, and only one-third of SERP pair relativities are
at the discretion of the individual metric.
At $k=20$ that fraction is still around half; and even at $k=50$
around one third of all SERP-pair relativities can still be regarded
as being metric-independent, and determined by application of the
Rule~1 and Rule~2 that were introduced earlier.

\myparagraph{Separability In Practice}
\vspace{-2mm}

\begin{table}[t]
\centering
\caption{Fraction of SERP pairs of length $k$ that are equal,
separable, and non-separable, based on all
SERP-vs-SERP comparisons in the TREC Robust Track data across $249$
  topics and $5{,}995$ system pairs.
\label{tbl-robpairs}}
\newcommand{\tabent}[1]{\makebox[12mm][c]{#1}}
\renewcommand{\tabcolsep}{3mm}
\begin{tabular}{cccc}
\toprule
$k$
	& \tabent{\sf{equal}}
		& \tabent{\sf{separable}}
			& \tabent{\sf{non-sep.}}
\\
\midrule
5
	& 20.40\%
		& 74.49\%
			& 5.12\%
\\
10
	& 8.33\%
		& 74.66\%
			& 17.01\%
\\
15
	& 4.91\%
		& 69.91\%
			& 25.18\%
\\
20
	& 3.55\%
		& 66.01\%
			& 30.44\%
\\
50
	& 1.54\%
		& 55.88\%
			& 42.58\%
\\
100
	& 0.92\%
		& 50.44\%
			& 48.64\%
\\
\bottomrule
\end{tabular}

 \aftertabspace
\end{table}

Table~\ref{tbl-robpairs} shows what happens when actual SERPs are
considered.
To build the table, all
$110$ systems submitted to the {\Robust} track were
compared to each other, over all $249$ of the track topics, with all
runs truncated at $100$ documents each.
The results aggregate the $249\times110\times109/2=1{,}492{,}755$
resulting SERP-vs-SERP relationships.
Compared to Table~\ref{tbl-allpairs}, a greater fraction of SERP
pairs are identical, and a greater fraction of the SERP pairs are
separable.
The lower-than-expected non-separability levels suggest that these
actual SERPs are not equivalent to being random.
Non-randomness arises in two quite different ways.
At low values of $k$ the proportion of {\Lab{1}}s and {\Lab{0}}s is
approximately equal (for example, at $k=5$ there are $50.5$\%
non-relevant and $49.5$\% relevant on average across the runs), but
the placement of those outcomes within the SERPs tends to be
correlated, as evidenced by the high number of ``equal'' outcomes in
the first row in Table~\ref{tbl-robpairs}.
The second effect is that as $k$ becomes larger there are far more
non-relevant documents than relevant (for example, when $k=50$ the
mix is $74.4$\% {\Lab{0}}s and $25.6$\% {\Lab{1}}s), caused in part
by the systems seeking to bring relevant documents to the beginning
of their runs, and then amplified by our application of the standard
IR assumption that unjudged documents are not relevant.

To quantify the extent to which unjudged documents might be a
confound, we explored the relevance judgments.
Across the set of $110 \times 259 = 27{,}390$ system-query runs,
$18{,}817$ ($68.7$\%) contained one or more unjudged documents, with
the average position of the first unjudged document across those
$18{,}817$ runs being rank $46.0$.
Within that set of $18{,}817$ runs there were on average $14.6$
unjudged documents per $100$-item run.
These numbers suggest that any measurements on the {\Robust} runs
that are taken at $k=100$ should be viewed with care, but that
$k\le20$ measurements can be considered to be reasonably reliable.
The appropriateness of using incomplete relevance judgments has been
an ongoing theme of IR research for more than two decades, and we
note that issue here without seeking to resolve it.

Regardless of the reasons, the fact that there are relatively high
separability levels in common system-vs-system experimental settings
at typical evaluation depths is an important outcome, and answers
{\bf{RQ2}}: a substantial fraction of paired SERP comparisons can
indeed be ordered using nothing other than their innate relationship
in the partial order that governs all SERPs.
Furthermore, these SERP pairs have been confirmed to occur in
practice {\cite{jtss15sigir}}.

\myparagraph{System Comparisons Over a Set of Topics}

The fact that a high fraction of the SERP-vs-SERP pairs that arise in
practice have innate orderings evident at plausible evaluation
depths~$k$ is very encouraging, and leads us to consider what happens
across sets of topics.

\begin{figure}[t!]
\centering
\includegraphics[width=120mm,clip,trim=15mm 180mm 15mm 5mm]{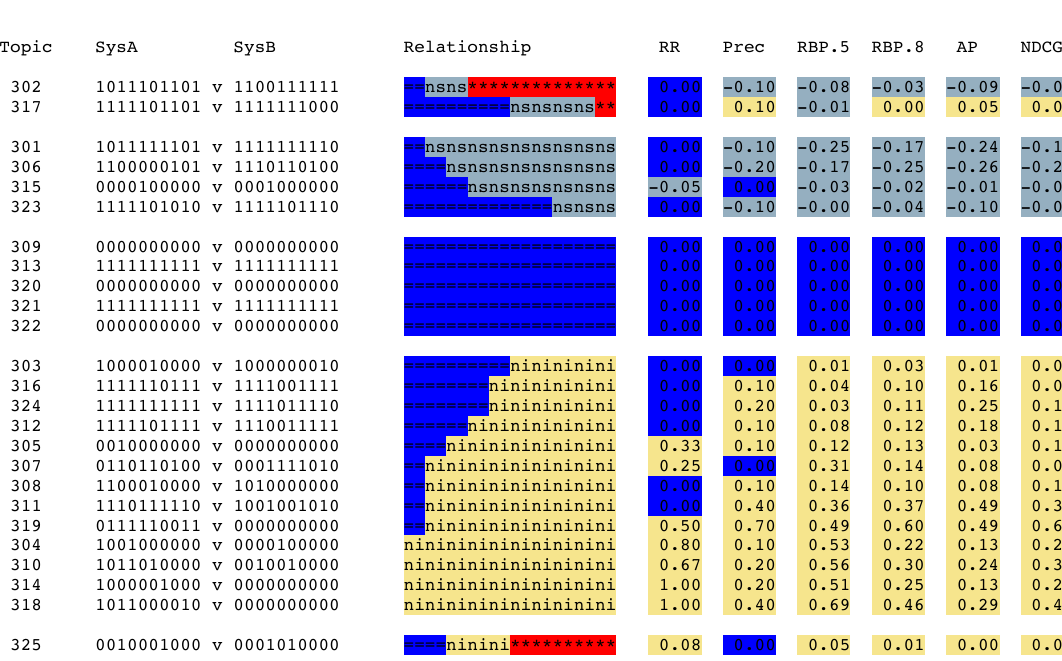}
\caption{Comparison to depth $k=10$ between two {\Robust} runs over
Topics $301$--$325$.
In the final six columns dark blue entries indicate equality of
metric scores; yellow entries indicate that System~$A$ scores more
highly; and light blue values indicate that System~$B$ scores more
highly.
The values in those cells are the score differences, $M(A)-M(B)$.
\label{fig-topics301to325}}
\end{figure}

Figure~\ref{fig-topics301to325} illustrates such a situation, taking
a set of $25$ topics ($301$--$325$) and two of the {\Robust} runs to
depth $k=10$, denoted System $A$ and System~$B$.
Each row shows a topic number; the two SERPs being compared, one from
System $A$ and one from System $B$; their {\IPSO} relationship using
the same labels and colors as were used in Figures~\ref{fig-rbp3}
and~\ref{fig-compare}; and then the metric score differences between
System~$A$ and System~$B$ for that topic for each of six different
effectiveness metrics.
(Note that the metric scores are not shown, only their signed
difference.)

At the same time the rows (topics) are ordered vertically into five
sections: the first group contains all the non-separable topics {\NP}
that include an {\NS} midpoint; the second group is all of
the separable topics that lead to an {\NS} outcome; the third group
consists of all of the {\EQ} topics; the fourth group shows the
separable topics that lead to an {\NI} outcome; and then the last
group shows the single {\NP} topic that has an {\NI}
midpoint.
Within each of the five groups the rows are ordered by the number of
red {\NP} cells (if any), and then by the extent of the leading blue
{\EQ} zone.
Given this overall ordering the rows can be thought of as ``wrapping
around'' so that the bottom row is also adjacent to the top; making a
cylinder that in fact contains four distinct sections rather than
five.

Looking at the metric score differences shown in the final six
columns of the figure, the ``$\le$'' requirement when System~$A$ and
System~$B$ are compared is clear in the second group of topics
(starting at Topic~$301$); and similarly the ``$\ge$'' requirement is
clear in the fourth group of topics (starting at Topic~$303$).
On the other hand, the top group and bottom group have no such
requirement, and in the top group (starting at Topic~$302$) all three
score difference possibilities can be seen, as was also noted in
connection with Table~\ref{tbl-k3-metrics}.
The one row that comprises the bottom group is consistent across the
six illustrated metrics, but that is just chance, and light-blue
{\NS} overall outcomes could arise if further metrics were added.

\begin{figure}[t!]
\center
\includegraphics[width=60mm]{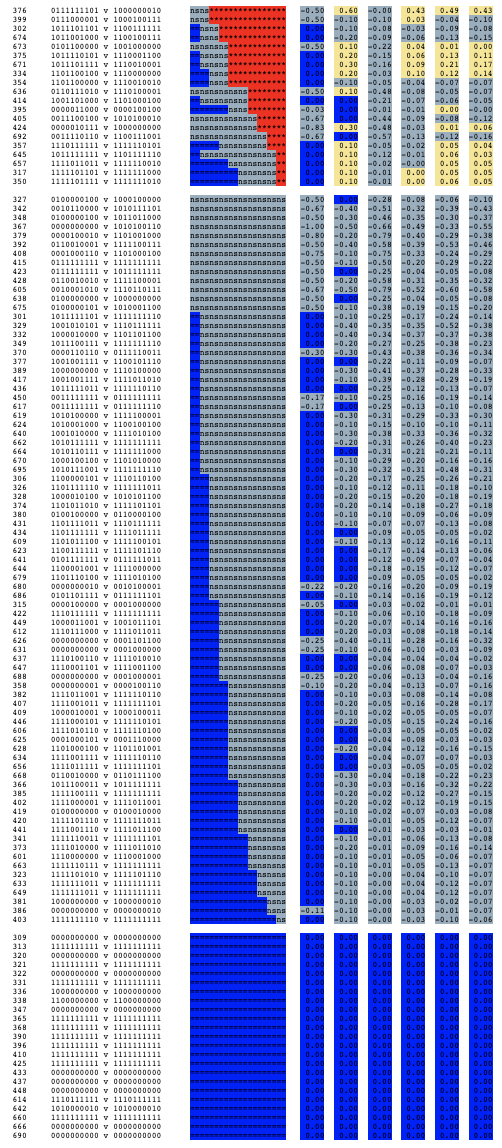}\makebox[3mm]{~}
\includegraphics[width=60mm]{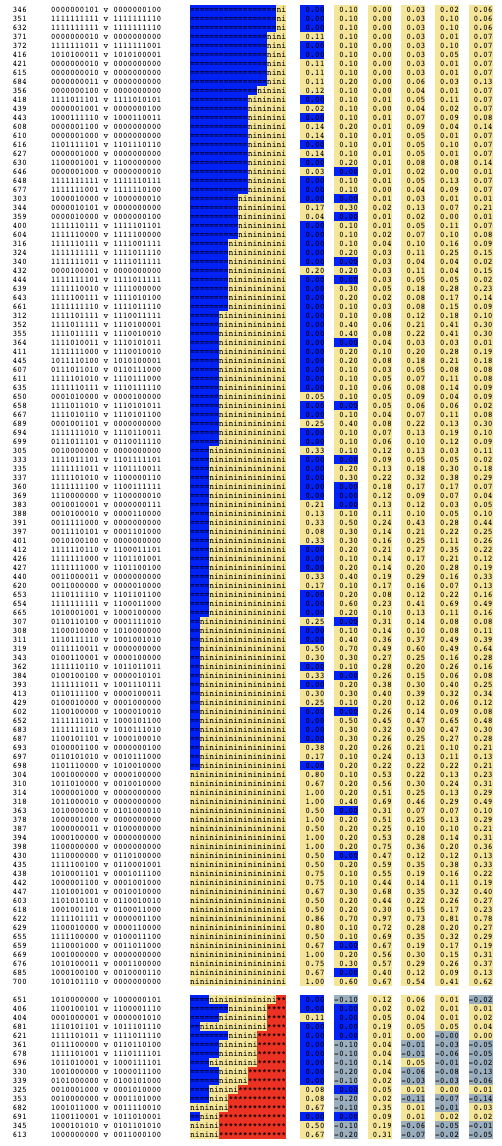}\caption{The {\IPSO} comparison over all $249$ {\Robust} topics to
depth $k=10$ between the same two systems shown in
Figure~\ref{fig-topics301to325}, using the same formatting and same
column headings, but here split into two pieces that are presented
side-by-side.
The five sections of the output contain $20$ topics, $81$ topics,
$23$ topics, $109$ topics, and $16$ topics respectively.
\label{fig-topics301to700}}
\afterfigspace
\end{figure}

Figure~\ref{fig-topics301to700} shows the same two {\Robust} systems,
but now with all $249$ topics included.\footnote{As with
Figure~\ref{fig-k6} we ask that the reader consider the patterns of
color, rather than squint to try and read the details of the SERPS
and of their metric score differences.}
There are again five distinct groups, or four when the top row is
considered to be ``adjacent'' to the bottom row, wrapped around a
cylinder.
Moreover, with $81$ topics in the second group, $23$ topics in the
middle group, and $109$ topics in the fourth group, it is clear that
for the great majority of topics the {\IPSO} analysis is sufficient
to identify the polarity of the paired SERP score differences.
Those mandated relationships are clearly visible for the six metrics
at the right of Figure~\ref{fig-topics301to700}, with columns for
$\RR$, ${\Prec}@10$, $\RBP(0.5)$, $\RBP(0.8)$, $\AP$, and $\NDCG$, as
in Figure~\ref{fig-topics301to325}.
Only $36$ of the topics yield $k=10$ SERP pairs for which an
effectiveness metric is free to assign scores in either order.
In those two sections of Figure~\ref{fig-topics301to700} both
positive (yellow) and negative (light blue) metric score differences
do arise.
But in the long second and fourth sections there is an enforced
inequality relationship on the per-topic score differences that
{\emph{simply cannot be violated}}.

\myparagraph{Statistical Testing}

The penultimate step in most system-vs-system experimental
comparisons is to carry out a paired statistical test on the metric
score differences, to establish the degree of support for the null
hypothesis that the mean difference is zero.
(The ultimate step being, of course, to crystallize the findings in to
a research paper submission).
If there is only limited support for the null hypothesis (with
``limited'' often at a level of $p<0.05$) it is rejected, and the
difference between the two systems is deemed to be ``significant''.

The IR community is well-versed in these techniques, and now expects
statistical testing, in one way or another, as a matter of
routine {\citep{sakai16sigir, sakai16irj, ulh19sigir,
sac07cikm, ferrosanderson22wsdm}}.
The Student $t$ test is probably the most commonly used option, but
the Wilcoxon Signed-Rank test and the simpler Sign test can also be
used, especially if the distribution of paired score differences does
not match the requirements for the parametric Student test (such as
when the metric being used is {\RR}).

Application of the {\IPSO} mechanism generates a set of five category
counts, but not scores.
Nor are there any score differences.
That means it is not possible to compute a $p$ value from the Student
$t$ test or from the Wilcoxon Signed-Rank test.
However, it is possible to employ a Sign test, because it applies to
{\emph{categories}} rather than to {\emph{values}}.
As an example, consider the system-vs-system relationship depicted in
Figure~\ref{fig-topics301to700}.
If the first and last groups of topics (of the five groups) are
discarded as being inconclusive, and the middle group is discarded as
being uninformative, what remains is a set of $81$ {\NS} topics and a
set of $109$ {\NI} topics.
It is then possible to pose as a null hypothesis that ``the number of
{\NS} topics is equal to the number of {\NI} topics'', and compute a
$p$ value relative to that hypothesis using the Sign text.
For the two illustrated runs that results in $p=0.0499$ as a
two-tailed outcome, and we can conclude that the difference between
$|\NS|=81$ and $|\NI|=109$ is significant at the $0.05$ level.
Note that it is the counts of the {\NI} and {\NS} topic sets that
contribute to the test, without reference to the overall topic set
size.
That way, if the {\NI} and {\NS} sets are closer to each other in
size, or if the two non-separable groups span the majority of the
topics and the {\NI} and {\NS} sets are small, the power available to
the Sign test will also be small, and the calculated $p$ values will
tend to be larger.

The $p$ values computed in this manner
must not be extrapolated as applying to the overall system comparison
for other metrics; to do so would be incorrect.
This is because in a normal score-based statistical test across a
full set of topics the queries are taken to be independent draws
from the population of all possible queries.
A small $p$ value then supports predictions that further independent
draws from the same underlying population will demonstrate the
same relationship between the two systems.
But in the scenario considered here the set of {\NP} topics across
which any such extrapolation would need to apply cannot be regarded
as being independent draws, and there is a selection bias.
Indeed, they are quite explicitly known to not have the same
characteristics as the ones that led to the {\NS} and {\NI} sets.
Moreover, as a further complexity, note that those two key classes,
{\NS} and {\NI}, are ``not superior'' and ``not inferior'', with both
including the possibility of metric score equality.
That is, a system comparison could, conceivably, assign every single
topic to the {\NS} set (or, equally, the {\NI} set) and thereby
attain a miniscule {\IPSO}-based Sign test $p$ value, and
yet for a particular metric, still have System~$A$ and System~$B$
assigned identical scores for every one of those topics.

Taking again the specific example shown in
Figure~\ref{fig-topics301to700}, there are $|\NP|=36$ queries for
which each effectiveness metric is free to assign numeric scores in
either relative order, and any {\IPSO}-based claims that are made
must of necessity respect that freedom.
Hence, depending on the metric, the final balance of ``votes''
between System~$A$ and System~$B$ might be anywhere between $81$
``versus'' $109+36$ (which yields $p<0.01$) at one extreme, and
$81+36$ versus $109$ (which results in $p\approx0.25$) at the other.

That is, the Sign test $p$ value generated from an {\IPSO}-based
analysis and the null hypothesis $|\NS|=|\NI|$ may be regarded as
being an interesting indicator of overall system relativities, but
must not be taken as a measurement of statistical confidence that
makes promises in regard to all metrics.

\begin{figure}[t!]
\centering
\includegraphics[width=130mm]{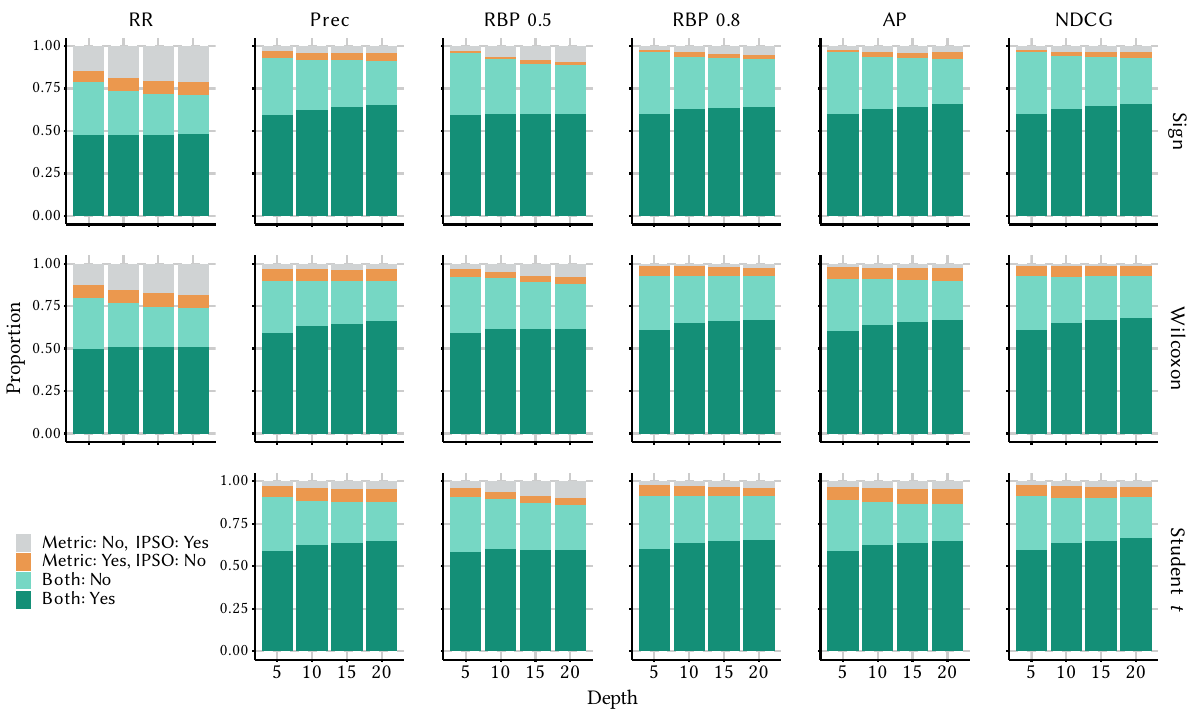}\\
\caption{Statistical test outcomes for all $110\times109/2=5{,}995$
System~$A$ versus System~$B$ challenger-vs-champion experiments
possible within the $110$ submitted {\Robust} runs.
In each case all $249$ topics are used, with six effectiveness
metrics then employed at each of four different retrieval depths;
three statistical tests (two-tailed in each case); and a significance
threshold of $p<0.05$.
The two-tailed Sign test is always used to obtain the {\IPSO} $p$
values, and they refer to the null hypothesis $|\NS|=|\NI|$.
\label{fig-hist2}}
\end{figure}

Given this context, Figure~\ref{fig-hist2} explores the relationship
between {\IPSO}-derived Sign test $p$ values (for the
null hypothesis $|\NS|=|\NI|$) and those generated via a range of
metrics (for the null hypothesis that
the two systems have the same performance as assessed by that
metric) in conjunction with three different statistical tests.
Each bar in each pane of the figure takes all possible System~$A$
versus System~$B$ pairs in the {\Robust} data, and applies one
effectiveness metric to one depth $k$, and then applies one selected
statistical test to the set of paired score differences, to compute a
$p$ value.
Each of those $(A, B)$ system pairs in each bar of each subgraph is
then labeled as being one of four possible categories:
\begin{itemize}
\item
``Both: Yes'', indicating that the {\IPSO} Sign test
(comparing $|\NS|$ and $|\NI|$) and the selected
metric/test pair
(comparing System~$A$ and System~$B$)
agree that there {\emph{is}} a significant
difference;

\item
``Both: No'', when the {\IPSO} Sign test and the selected metric/test
pair agree that there is {\emph{not}} a significant difference;

\item
``Metric: Yes'', where the metric/test combination reports
significance, but {\IPSO} does not;
and, finally 

\item
``Metric: No'', where {\IPSO} reports significance, but the
metric/test combination between System~$A$ and System~$B$ does not.
\end{itemize}

Of these four possibilities, the combined ``Both: Yes'' and ``Both:
No'' categories represent a reassuringly high level of agreement.
They are shown in Figure~\ref{fig-hist2} as two shades of green at
the bottom of each of the bars in each of the panes, and cover a
minimum (across the bars and panes) of three quarters of all system
pairs, and up to $90$\% or more of the system pairs in some
metric/test cases.
Note also how (with the exception of $\RR$) the fraction of ``Both:
Yes'' tends to increase with $k$ regardless of metric and statistical
test; and how for any given metric the increased power of the
Wilcoxon and then $t$ tests relative to the Sign test is visible via
the growth of the mustard-colored ``Metric: Yes'' category when
moving down each column of graphs.

\begin{figure}[t!]
\centering
\renewcommand{\tabcolsep}{1em}
\begin{tabular}{@{}cc@{}}
\includegraphics[width=60mm]{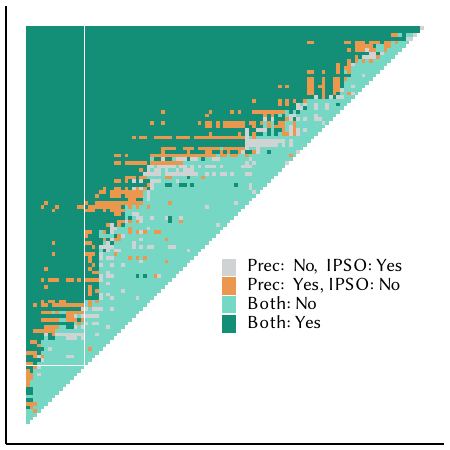}
&
\includegraphics[width=60mm]{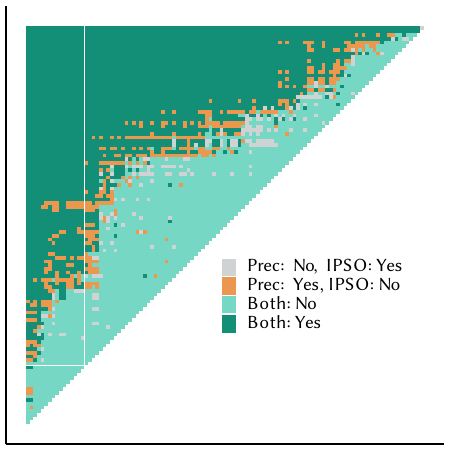}
\\
(a) $249$ topics, $p<0.05$
&
(b) $249$ topics, $p<0.01$
\\
\includegraphics[width=60mm]{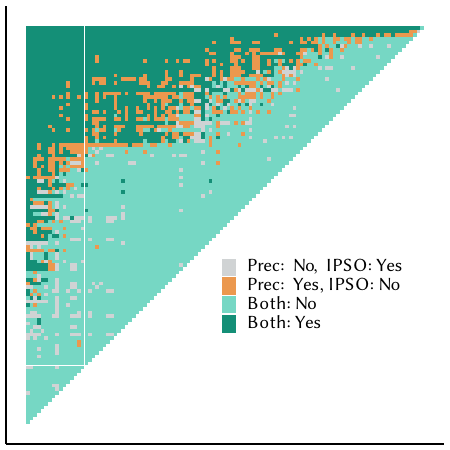}
&
\includegraphics[width=60mm]{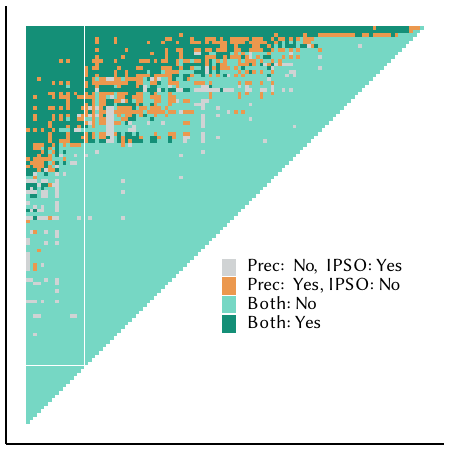}
\\
(c) $50$ topics, $p<0.05$
&
(d) $50$ topics, $p<0.01$
\\
\end{tabular}
\caption{System-vs-system significance categories, with one ``dot''
plotted per system pair using colors that match
Figure~\ref{fig-hist2}.
The ordering of the $110$ {\Robust} systems is based on their average
${\Prec}@10$ score across the given topic set (all $249$ topics in the
top two panes, and topics $301$--$350$ in the bottom two panes), and with
system-vs-system significance calculated from paired ${\Prec}@10$
scores using the Student $t$ test.
\label{fig-heat}}
\end{figure}

The non-green minority of cases represent system pairs for which
disagreement occurs.
Figure~\ref{fig-heat} helps understand the situations in which that
happens.\footnote{{\citet{vss17tois}}, and perhaps others, make
use of a similar triangular presentation.}
To form each pane the $110$
systems submitted to the {\Robust} track were ordered by
average ${\Prec}@10$ score, and then the $5{,}995$ individual
system-vs-system experiments that were aggregated into the
``${\Prec}@10$, $k=10$, Student $t$ test'' bar in
Figure~\ref{fig-hist2} were each represented by a colored dot placed
according to the ranks of the two systems.
For example, the dot at bottom left in each pane represents the
``best'' versus ``second best'' system, and the dot at top right is
``second worst'' versus ``worst''.
Dots along the diagonal edge similarly represent systems being
compared to the ones immediately adjacent to them in the system
ordering induced by ${\Prec}@10$.

In each of the four panes in Figure~\ref{fig-heat} there is a
distinctive pattern of dark green (both ${\Prec}@10$ and $\IPSO$
indicate significance) and light green (neither indicate
significance), with a transition region between of grey and
mustard-colored points.
Note how increasing the stringency of the significance threshold
(moving from pane~(a) to pane~(b), and similarly moving from pane~(c)
to pane~(d)) and reducing the number of topics employed (moving from
pane~(a) to pane~(c), and similarly moving from pane~(b) to pane~(d))
both shift the balance between dark green and light green, but don't
alter the overall pattern.
Note also how the grey ``$\IPSO$ only'' cells tend to spread into
(that is, be surrounded by) the light green zone, whereas the
mustard-colored ``${\Prec}@10$ only'' cells tend to spread into the
dark green zone.
In Figure~\ref{fig-heat}(a), $67.4$\% of the plotted dots are dark
green and $7.5$\% are mustard, with ${\Prec}@10$ identifying a total of
$74.9$\% of system pairs as being significantly different when the
Student $t$ test is employed.
On the other hand, $\IPSO$ and the Sign test achieve a total of
$71.4$\%.
That these rates are different is of no concern.
In particular, the statistical tests are different, and $\IPSO$
significance is in one sense a broader outcome than is ${\Prec}@10$
significance, making it harder to achieve, but at the same time is
also a weaker outcome, since it is based only on the $\NS$ and $\NI$
counts (with both groups permitting score equality), rather than on
score differences.

The two systems depicted in Figures~\ref{fig-topics301to325}
and~\ref{fig-topics301to700} are in fact the best {\Robust} run and
the $25$\,th percentile run according to ${\Prec}@10$ -- which, as a
pair, correspond to points 1/4 of the way up the triangle's left edge
in each pane of Figure~\ref{fig-heat}.
Direct metric-based comparisons of these two runs can also be carried
out, of course.
For example, ${\Prec}@10$ yields $p=0.017$ using a two-tailed Student
$t$ test, and $\NDCG$ gives $p=0.027$.
On the other hand, $\AP$, $\RBP(0.5)$, and $\RBP(0.8)$ all result in
$p\geq0.05$.
This kind of variability is inevitable when statistical tests are
employed to gauge significance, and the reassurance, or ``hedge'' as
it was described above, is provided by noting that all six metrics
assess the first run as being better than the second.
Such differences in system performance can exist without a
statistical test passing some selected -- but also arbitrary --
threshold.
The critical point that we seek to make in this work is that the
{\IPSO}-based Sign test has the capacity to encompass all of those
metric-specific outcomes and provide
overall guidance as to the likely ordering of the two systems.

The results we have presented thus answer {\bf{RQ3}}:
{\IPSO}-derived system-vs-system relationships generally agree with
those of conventional effectiveness metrics under commonly applied
statistical tests.

\myparagraph{Using IPSO as a Corroboration}

We have now arrived at our proposal.
As was noted earlier, it is common for researchers reporting the
result of a system-vs-system comparison to tabulate several metrics,
and a statistical test for each, as a way of providing multiple
sources of evidence that their new challenger system does indeed
outperform the current champion.
We propose instead that the hedging aspect of multi-metric analysis
be supported by an {\IPSO}-based system~comparison.

For example, suppose that two systems $A$ (champion) and $B$
(challenger) are being compared in regard to their suitability for
some defined search task.
We suggest that the comparison be carried out as follows:
\begin{itemize}

\item
Select a {\emph{single}} evaluation metric $M(\cdot)$ for which the
corresponding user model can be argued as matching the anticipated
(or observed) behavior of searchers when carrying out that task, and
the way that particular user population might therefore define and
measure ``search success''; 

\item
Report mean metric scores $M(A)$ and ${M}(B)$
and the mean score difference $M(B)-M(A)$ (the measured effect size);

\item
Then carry out a suitable significance test against the null hypothesis that
``the effect size is zero'';

\item
If that test yields a $p$ value less than some stipulated threshold
$\alpha$ (typically $\alpha=0.05$), then ``add a $\dagger$'' and claim
significance;

\item
Plus, if the metric-based $p$ value is less than $\alpha$, perform an
{\IPSO}-based Sign test against the null hypothesis that ``the number
of {\NS} topics and the number of {\NI} topics is the same'';

\item
And if $p<\alpha$ from that second test as well, add that further
claim (by adding a $\ddagger$, or some other preferred symbol) to the
reported results, as a corroboration of generality.
\end{itemize}
That is, we argue in response to {\bf{RQ4}} that one {\IPSO}-based
test ``in the hand'' is worth multiple other metrics ``in the bush''
when seeking to add generality to an evaluation that has, for
principled reasons, been based on a chosen user-matched effectiveness
metric.

\myparagraph{Implications and Limitations}

If a single pair of SERPs lead to an {\IPSO} outcome of $\NI$ or
$\NS$, then the relationship between the two SERPs in an ``at $k$''
comparison is certain and is metric-agnostic.
On the other hand, as has been noted above, if an
{\IPSO}-based Sign test over a set of topics yields a significant
outcome, there is still no confidence that the outcome is applicable
to all of the tested topics, nor to all future topics.
That lack of confidence is partly a consequence of the nature of
statistical testing -- it gives encouragement (based on a
representative sample) but never certainty;
and is partly a consequence of the fact
that {\IPSO} is an imprecise indicator, with the {\NS} and {\NI}
classes both admitting score equality, and with the {\NP} class able
to swing either way.

Also worth noting is that -- as is also the case for standard metric
scores -- each SERP length $k$ is likely to lead to a different
computed $p$ value.
In the case of most metrics, as the evaluation depth increases, so
too does the likelihood of significance, an effect that was
identified in connection with Figure~\ref{fig-hist2}.
But in the case of {\IPSO}, the fact that the $\NI$ and $\NS$ topic
counts in any given set cannot increase as $k$ increases means that
the $p$ value will, other things being equal, also tend to increase.
That is, the greater the value of $k$, the less likely will be a Sign
test $p$ value that is below the threshold.
We will undertake further experimentation to explore this aspect of
our proposal.

\myparagraph{Graded Relevance}

The examples used as illustrations in
the previous section all made use of binary relevance labels; as did
the collections and runs used in the experiments presented in this
section.
But exactly the same definitions (Rule~1 and Rule~2) can be applied
to real-valued (ratio scale) SERP
comparisons, without any adjustments being required.
Similarly,
Algorithm~\ref{alg-compare} remains the correct description of the
comparison computation, except that {\Lab{S1}} and {\Lab{S2}} are now
vectors of floating point values.
For example, if relevance is being reported on a four-point
ratio\footnote{Meaning that relevance fractions are additive across
documents, so that a $0.2$ document in conjunction with a $0.8$
document is exactly as useful as a single $1.0$ document, for
example.}
scale, $r_i\in\{0.0,0.2,0.8,1.0\}$ say, and we have $k=5$, then
$\Lab{S1} = [1.0,0.8,0.0,0.2,1.0]$ and
$\Lab{S2}=[0.8,0.8,0.0,0.2,0.8]$ have the relationship
$\Lab{S1}\succeq\Lab{S2}$ because of Rule~1; and if $\Lab{S3} =
[1.0,0.2,0.0,0.8,1.0]$ then $\Lab{S1}\succeq\Lab{S3}$ because of
Rule~2; and $\Lab{S2}$ and $\Lab{S3}$ are non-separable.

If the relevance grades are ordinal classes such as ``not at all'',
``somewhat'', and ``highly'', then a gain mapping that converts
ordinal relevance grades to floating point values must be argued for
and applied first, in exactly the way as is needed for all numeric
score comparisons based on graded relevance evaluations.
That is, Algorithm~\ref{alg-compare} may not be applied directly to
relevance grades such as $r_i\in \{0,1,2\}$ representing
$r_i\in\{\var{not at all}, \var{somewhat}, \var{highly}\}$, because
the computation in Algorithm~\ref{alg-compare} requires addition, and
ordinal class labels such as ``somewhat'' may not be summed, even
when it represented by the faux-integer value of~``1''.
That is, the ordinal-to-numeric gain mapping
{\citep{ieeeaccess22moffat}} must be specified before the scoring can
be considered, and while {\IPSO} provides metric-independent
assessments, it cannot be gain-mapping-independent.

Experimental evaluations in graded-relevance
measurement contexts will be carried out as future work.

 \section{Related Work}

The most pertinent prior work is that of {\citet{df22sigir}}.
Their {\emph{recall paired preference}} (\method{RPP}) approach also
carries out SERP comparisons ``without effectiveness metrics''.
In this mechanism a distribution over users is assumed as to how many
relevant items are being sought from the SERPs, and for each user the
SERP that delivers that many relevant documents more quickly is the
one that is preferred.
As a distinction to previous work, {\citeauthor{df22sigir}} note
that, like {\AP}, their probability distribution is over recall
levels rather than rank positions, differentiating from metrics such
as {\RBP} and {\method{DCG}}.
The relationship between two SERPs is then determined by forming a
preference expectation over the defined user population.
Like our {\IPSO} method, {\method{RPP}} reports an ordering between
two SERPs rather than scores for single SERPs that must then be
compared.
In the case of {\citeauthor{df22sigir}}, the distribution of user
goals informs the outcome and a SERP pair always leads to a $<$, $=$,
or $>$ outcome.
In our work here we assume that users look at $k$ or fewer documents,
seeking generality that covers all $@k$ metrics, once the validity of
Rules~1 and~2 is accepted; but thus also needing to allow ``don't
know'' answers (the non-separable pairs) as well as $\leq$, $=$, and
$\geq$ determinations.
{\citet{diaz23arxiv}} and {\citet{dm23arxiv}} go on
to consider two further formulations to inform our understanding of
how to compare SERPs with each other: {\method{lexiprecision}} and
{\method{lexirecall}}, developed by considering the lexicographic
properties of SERPs.
These methods compare SERPs based on the position in the ranking of
the first relevant item that is not matched by relevance in the other
SERP, or the rank position of the $p$\,th relevant item for some
selected value $p$, or the rank position of the last relevant item,
or some amalgam of such values.

In earlier work that anticipates our proposal here,
{\citet{jtss15sigir}} note that there is only a subset of depth-$k$
rankings on which metrics can disagree.
They then provide exhaustive experimentation on ranked results lists
of depth $k = 10$, with a total of $10$ relevant documents available
(resulting in $1{,}023$ such lists when ignoring the ``all zeros''
ranking), and confirm that all such rankings do occur in organic
ranking tasks, by empirically examining submitted TREC runs.
They also provide a more detailed analysis on when metrics tend to
disagree, and explore properties that may be predictive of such
disagreement.

The relationship between user behavior and the underlying quality of
ranked results lists has been explored by a range of authors,
resulting in a number of metrics for evaluating such rankings
{\citep{mz08-tois, robertson08sigir, cmzg09-cikm, rky10sigir}} and a
greater understanding of the relationship these metrics have with
assumed user models {\citep{sr08evia, cmzg09-cikm, dp10sigir,
dupret11spire, carterette11sigir, cky12cikm, mbst17acmtois, wm18cikm,
zlmzm20aiopen, sigir22mmta}}.
The overarching theme of all of these papers is that metric scores
should be formulated in the context of how users are likely to derive
information from SERPs, so as to allow the computed scores to reflect
an informative quantity; it is our strong desire to retain and
embrace that connection that has informed our proposal here.
In particular, it is important to note that we are {\emph{not}}
arguing for {\IPSO} evaluation to supersede metric-based evaluation;
rather, we suggest {\IPSO} as an augment for a well-chosen metric.

Ideally, a metric score should be reflective of user satisfaction, as
this is what ultimately matters when measuring ranked retrieval
systems.
To this end, a number of works have sought to establish a connection
between metric score and user satisfaction.
This then allows metrics to be, at least in part, validated against
the user experience
{\citep{
hh07sigir,
jiang16chiir,
mlz-etal16sigir,
czl-etal17sigir,
shl-etal18wsdm,
llm-etal18www,
oac20ipm,
zhang-etal20sigir,
sz21acmtois,
sigir22mmta}}.
On the other hand, {\citet{ahm22arxiv}} have recently noted use cases
in which it could be important to measure SERP differences that might
{\emph{not}} be discernible by users, such as in diagnostic
situations, or perhaps as a model training objective.

{\citet{ffp17ictir,fff21ieeeaccess}} have presented arguments to the
effect that many traditional effectiveness metrics may not be
well-founded from a measurement point of view because they are not on
an interval scale; a claim that is strongly contested
{\citep{sakai20forum,ieeeaccess22moffat}}.
Irrespective of the outcome of that debate, {\IPSO} presents a
metric-free way of carrying out system comparisons to obtain guidance
as to which system might be preferable.
Indeed, this is one inherent benefit of the Sign test, which does not
require magnitudes of differences for meaningful hypothesis testing.

{\citet{moffat13airs}} categorizes metrics according to seven numeric
properties, including some that embed Rules~1 and~2; and a number of
other researchers have also considered measuring effectiveness from
what is referred to as an ``axiomatic'' viewpoint, of which Rules~1
and~2 are instances {\citep{bollmann84sigir, bm13ictir, afmz17sigir,
gkkl19ecir, afmz20irj, g22ictir}}.
Axiomatic approaches have also been used in the quest for retrieval
models {\citep{fz05sigir}}.

The process of planning and executing experimental system comparisons
has also received considerable attention, including in regard to
topic set size and statistical testing practices {\citep{sac07cikm,
sakai16irj, sakai16sigir, ulh19sigir, ferrosanderson22wsdm}}.
{\citet{sakai16irj}} considers the design of batch retrieval
experimentation in terms of the number of topics to employ to reach
statistically valid outcomes.
{\citet{sakai16sigir}} also provides a detailed systematic review on
testing practices across over $1000$ {\emph{SIGIR}} and {\emph{TOIS}}
papers from 2006 to 2015.
Other work has explored the reliability and predictivity of typical
batch IR experiments and measurement protocols {\citep{cl06-sigir,
rzm21-sigir}} and some of the limitations therein~{\citep{forum23zobel}}.

Finally, work that describes the overall process of batch evaluation
and issues that arise in the formation of relevance judgments may
also be of interest to the reader {\citep{saracevic95sigir,
voorhees02clef, bv04sigir, bv05trecbook, sn08irj, s10-fntir,
hofmann2016fntir}}.

 \section{Conclusion}

We have described {\IPSO}, a mechanism for comparing two SERPs that
is based on fundamental ordering criteria that apply to every
plausible effectiveness metric.
While {\IPSO} is not itself a metric, and does not generate
effectiveness scores, it can nevertheless be used as input to the
Sign test and hence derive a $p$ value when two systems are being
compared over a set of topics.
That is, {\IPSO} can be used to provide evidence that a measured
relationship in a challenger-vs-champion experiment holds not just
for the chosen metric -- which for experimental fidelity reasons
should always be selected (and/or parameterized) based on the user
characteristics that are anticipated for that type of search -- but
(subject to the limitations we have
noted) for other metrics as well.
In particular, a System~$A$ versus System~$B$ comparison]using a
chosen metric $M(\cdot)$ in which $M(A)-M(B)<0$ with statistical
confidence that is accompanied by an {\IPSO}-based significance
outcome that also favors System $B$ is a stronger result than one
based solely on the metric alone, and an augmented
evaluation of the suggested type may be a more powerful demonstration
of generality than is possible with a table employing multiple
different effectiveness metrics.

The new {\IPSO} mechanism also has limitations, of course.
First is the need to understand that it is a comparison tool, not a
metric than can be used to score runs.
It does not provide a numeric value that tells you how good a SERP is
in isolation, and even when it says that one SERP is ``not inferior''
to another, it cannot quantify the extent of the possible
superiority.
Nor can it indicate how close to being ``perfect'' any particular run
is.
For those tasks, a metric is required, meaning that a specific user
model must also have been selected.
As noted earlier, care is also required
when interpreting the {\IPSO}-derived $p$ values.
Any significance observed relates to the null hypothesis
$|\NS|=|\NI|$, and not to the more targeted hypothesis that
``System~$A$ and System~$B$ have the same performance when measured
using metric $M(\cdot)$''.

Nor is {\IPSO} invariant in its evaluation with regard to depth $k$.
As can be seen in Figure~\ref{fig-compare}, any given {\IPSO}
evaluation is likely to start by indicating equality between the two
SERPs (this is the starting point for $k=0$, of course), then shift
to a period of either ``non-inferior'' or ``non-superior'', and then,
as $k$ gets even larger (and sufficiently deep judgment are available
to support continued evaluation), may well undertake a second
transition to ``non-separable''.
That is, an {\IPSO} Sign-test relationship established at one value
of $k$ might not continue to apply at a larger value of $k$.
We note that the same risk also applies to comparisons based on
metric scores -- a significant relationship established at one value
for $k$ might not be recognized as being significant at a different
value of $k$.
As a component of further experiments
to investigate more closely any sensitivity to the value of $k$, we
also need to carry out an extended experimental comparison across a
range additional datasets, including ones that employ
graded-relevance scoring regimes, in order to provide further
assurance that {\IPSO} leads to outcomes that can be usefully
interpreted.

Despite these limitations, we recommend that researchers incorporate
{\IPSO} into their experimental tool chains, and employ our suggested
reporting framework in which one metric is adopted, defended, and
then deployed, with {\IPSO} used as the ``hedge'' to provide evidence
in regard to generality.

Finally, note that the definition of
{\IPSO} is fundamentally tied to the assumption that users consume
each SERP sequentially.
We posit that to be a reasonable assumption for traditional linear
text/link-based SERPs (and note that many previous authors have also
employed the same assumption), but it may not apply to (for example)
the two-dimensional SERP presentations that emerge from image and
product search.
Each user of those services may have a regular browsing sequence that
they tend to follow, but different users might have different
inspection orderings.
Extending our analysis to this important case in a key area for
future work.
 
\myparagraph{Software}

Source code to allow {\IPSO}-based comparisons on
pairs of SERPs is provided at
{\url{https://github.com/JMMackenzie/IPSO}}.

\myparagraph{Acknowledgements}

This work was in part supported by the Australian Research Council
(project DP190101113).
We thank the referees for their supportive and insightful comments.

\end{document}